\documentclass[10pt,journal,final,finalsubmission,twocolumn]{IEEEtran}
\usepackage{stmaryrd}
\usepackage{amsfonts}

\usepackage{mathrsfs,amsmath}
\usepackage{graphicx,times,amsmath}
\usepackage{cleveref}
\usepackage{cite}
\usepackage{authblk}
\usepackage{amssymb}
\usepackage{makeidx}
\usepackage{multirow}
\usepackage{subfigure}
\usepackage{algorithmic}
\usepackage{amsmath}
\usepackage{balance}
\IEEEoverridecommandlockouts    % to create the author's affliation portion
\renewcommand\thepage{}

\allowdisplaybreaks

\begin{document}

\title{Resource Allocation Technique for Hybrid TDMA-NOMA System with Opportunistic Time Assignment\thanks{The work of O. A. Dobre was supported by the Natural Sciences and Engineering Research Council of Canada (NSERC), through its Discovery Program.}}
\author{\IEEEauthorblockN{Xinchen~Wei$^{*}$, Haitham~Al-Obiedollah$^{\P}$, Kanapathippillai Cumanan$^{*}$, Miao~Zhang$^{*}$, Jie Tang$^{\dag}$, Wei Wang$^{\ddag}$ and Octavia~A.~Dobre$^{\S}$}\\
\IEEEauthorblockA{Department of Electronic Engineering, University of York, York, YO10 5DD, United Kingdom$^{*}$\\
The Hashemite University, Jordan$^{\P}$\\
School of Electronic and Information Engineering, South China University of Technology, Guangzhou, China$^{\dag}$\\
School of Information Science and Technology, Nantong University, Nantong 226000, China$^{\ddag}$\\
Research Center of Networks and Communications, Peng Cheng Laboratory, Shenzhen, China$^{\ddag}$\\
Department of Electrical and Computer Engineering, Memorial University, St. John's, Canada$^{\S}$}\\
Email: \{xw1647, kanapathippillai.cumanan, mz1022\}@york.ac.uk$^{*}$, haithamm@hu.edu.jo$^{\P}$, eejtang@scut.edu.cn$^{\dag}$, wwang2011@ntu.edu.cn$^{\ddag}$, odobre@mun.ca$^{\S}$.
\vspace{-10mm}}
\maketitle
% reference style
\bibliographystyle{IEEEtranN}
\begin{abstract}
In this paper, we develop a resource allocation technique for a hybrid time division multiple access (TDMA) - non-orthogonal multiple access (NOMA) system with opportunistic time assignment. In particular, the available transmission time is divided into several time-slots, through which multiple users are served by exploiting power-domain NOMA. To fully exploit underlying benefits of this hybrid TDMA-NOMA system, we utilize the available resources efficiently by jointly allocating transmit power and time-slots to several groups of users in the system. Furthermore, these resources are allocated to maximize minimum rate of the users in the system. However, this max-min resource allocation problem is non-convex due to coupled design parameters of time and power allocations. Hence, we exploit a novel second-order cone formulation to overcome this non-convexity issue and develop an iterative algorithm to realize a solution to the original max-min problem. Simulation results show that this joint resource allocation technique has a considerable performance enhancement in terms of both minimum achieved rate and overall system throughput compared to that of the conventional resource allocation technique where equal time-slots are assigned to the groups of users.
\end{abstract}


\section{Introduction}

Non-orthogonal multiple access (NOMA) has been recently proposed as a promising multiple access technique for the fifth generation (5G) and beyond wireless networks due to its potential benefits including superior spectral efficiency and user fairness \cite{Y,alavi2}. Unlike the conventional orthogonal multiple access (OMA) schemes, such as time division multiple access (TDMA) and orthogonal frequency division multiple access (OFDMA), multiple users in NOMA based downlink transmission share the same orthogonal radio resources, i.e., time and frequency, by exploiting power-domain multiplexing at the transmitter \cite{Benjebbour}. This multiplexing is referred to as superposition coding (SC), in which signals intended to different users are encoded with different power levels that are inversely proportional to the channel strengths of the users \cite{Demestichas,Alavi}. In particular, serving multiple users simultaneously within the same resource block through NOMA supports the proliferation on Internet-of-Things (IoTs) by offering massive connectivity \cite{Rappaport}. At the receiver end, successive interference cancellation (SIC) technique is utilized at stronger users to decode the signals intended to the weaker users prior to decoding their own signals \cite{Islam,Higuchi,10,11}. However, the computational complexity of SIC increases as the number of served users increases in the system \cite{phd,alavi1}. 

To overcome the practical challenges of employing SIC in dense networks, and to meet the unprecedented requirements of future wireless networks, NOMA has been integrated with different other technologies, including multiple-antenna techniques \cite{Kan,Timotheou,Jingjing,O.A.Dobre} and conventional OMA schemes \cite{M,H,Elkashlan,Fang,Wei,Wei2,Diamantoulakis}. In a hybrid OMA-NOMA systems, the available resources (i.e., time or frequency) are divided into several sub-resource blocks and each sub-resource block is assigned to serve multiple users based on NOMA \cite{H,Fang}. For example, a hybrid TDMA-NOMA system has been considered in \cite{H}, in which the available time for transmission is divided equally among several groups of users (i.e., clusters), and the energy harvesting capabilities of such system is investigated. A hybrid OFDMA-NOMA system is considered in \cite{Fang}, where the available bandwidth is divided into several sub-bandwidths, and the available resources are allocated to maximize the energy efficiency of the system. In \cite{Wei} and \cite{Wei2}, different resource allocation techniques for hybrid OFDMA-NOMA systems are developed. In fact, these combinations not only simplify the implementation of SIC, but also offers additional degrees of freedom by utilizing different domains to serve multiple users. Considering the hybrid TDMA-NOMA system, the work in the literature assume equal time assignments to serve the available groups of users to reduce computational burden at the receiver ends. However, this equal time assignments limit the performance enhancement of such a hybrid TDMA-NOMA system owing the fact that opportunistic time allocations provides additional benefits to the groups of users. Furthermore, serving each user in such hybrid TDMA-NOMA systems to achieve reasonable throughput is one of the key objectives of such systems. However, maximizing overall throughput of the system degrades the performance of individual users while compromising user-fairness in terms of achievable rates.  

Motivated by these facts, we aim to maximize the minimum per user rate for a single-input single-output (SISO) hybrid TDMA-NOMA system, while satisfying the relevant constraints on the system. This can be achieved through developing an optimization framework to allocate the available transmit power efficiently among users and opportunistically assign the available time for transmission between the clusters (i.e., groups of users). The formulated optimization problem is non-convex in nature, and cannot be solved via available software. Hence, we develop an iterative algorithm by exploiting sequential convex approximation (SCA) \cite{Beck}. Furthermore, a novel form of a second-order cone (SOC) is utilized to cast some of non-convex constraints as SOCs. Finally, we draw a number of performance comparisons to demonstrate the advantages of the proposed hybrid NOMA-TDMA technique over the conventional schemes with equal time allocations.

The remainder of the paper is organized as follows. Section II describes the system model of a hybrid TDMA-NOMA system and formulates the max-min resource allocation problem. In Section III, the developed joint resource allocation technique is presented with technical details, whereas Section IV provides simulation results and discussions. Finally, Section V concludes the paper. 

\section{System Model And Problem Formulation} \label{Sec:Max-minSINR}
\subsection{System Model}

We consider a downlink transmission of a multi-user SISO hybrid TDMA-NOMA system, as shown in Fig. 1. In this hybrid system, a base station (BS) with single-antenna communicates with $K$ single-antenna users. All $K$ users are grouped into $C$ clusters and a time-slot $t_i, \forall i ={1,2,...,C}$, is assigned to serve the $i$th cluster ($G_i$). In particular, the available transmission time ($T$) is divided into $C$ sub-time slots, as shown in Fig. 2, such that $T=\sum_{i=1}^{C}t_i$. Furthermore, the users at each sub-time slot are served based on power-domain NOMA technique.

\begin{figure}
\centering
\includegraphics[width=0.5\textwidth]{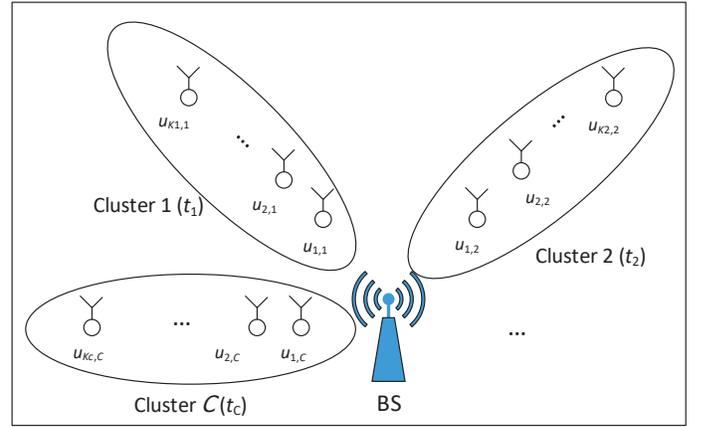}
\caption{A multi-user SISO hybrid TDMA-NOMA system.}
\label{fig_graph}
\end{figure}

\begin{figure}
\flushleft
\includegraphics[width=0.5\textwidth]{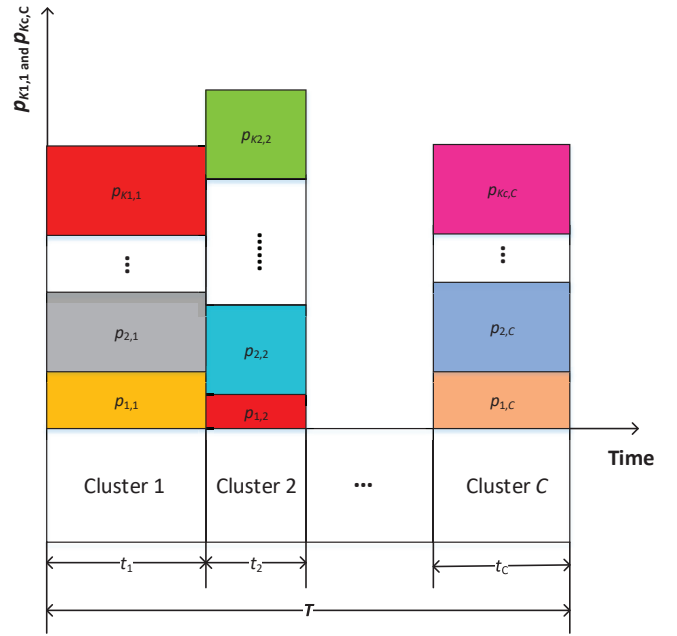}
\caption{A time-slot is assigned to serve each cluster, while the users in the corresponding cluster exploit power-domain NOMA to communicate with the BS.}
\label{fig_graph}
\vspace{-4mm}
\end{figure}
All $K$ users in the system are grouped into a number of clusters; the number of users in the $i$th cluster (i.e., $G_i$) is denoted by $K_i$, $\forall{i} \in \mathcal{C} \overset{\vartriangle}=\{1,2,...,C\}$ and $K=\sum_{i=1}^C{K_i}$. On the other hand, $u_{j,i}$ represents the $j$th user in the $i$th cluster. Note that user grouping strategy plays a crucial role in determining the performance of the system. Therefore, we shed some light on the proposed user grouping strategy later on this paper. 

Now, the transmitted signal from the BS during $t_i$ can be written as follows:
\begin{equation} 
x_i=\sum_{j=1}^{K_i}p_{j,i}x_{j,i},
\end{equation}
where $x_{j,i}$ and $p_{j,i}$ are the symbol intended to $u_{j,i}$ and the corresponding power allocation, respectively. Note that the channel coefficients between the BS and the user $u_{j,i}$ are denoted by $h_{j,i}$, $i=1,2,...,C$, $j=1,2,...,K_i$. Furthermore, it is assumed that the BS and the users have perfect channel state information. The received signal at $u_{j,i}$ can be defined as follows\cite{xinchen}:
\begin{equation} 
r_{j,i}=h_{j,i}x_i+n_{j,i}, \forall{i} \in \mathcal{C}, \forall{j} \in \mathcal{K}_i \overset{\vartriangle}= \{1,2,...,K_i\},
\end{equation}
where $n_{j,i}$ denotes the additive white Gaussian noise (AWGN) with zero-mean and variance $\sigma_{j,i}^2$ dBm/Hz at receiver.

Without loss of generality, it is assumed that the users at each cluster are ordered based on their channel strengths, such that,
\begin{equation} 
|h_{1,i}|^2 \geq |h_{2,i}|^2 \geq ... |h_{K_i,i}|^2, \forall{i} \in \mathcal{C}.
\end{equation}
Specifically, the SIC process is implemented at stronger users, i.e., user $u_{j,i}$ performs SIC to decode and remove the signals intended to the users from $u_{j+1,i}$ to $u_{K_i,i}$ prior to decoding its own signal, where it is assumed that SIC is implemented perfectly without any errors in this paper. Therefore, the received signal at $u_{j,i}$ after employing SIC can be expressed as
\begin{equation} 
r_{j,i}^{SIC}=h_{j,i}p_{j,i}x_{j,i}+h_{j,i}\sum_{s=1}^{j-1}p_{s,i}x_{s,i}+n_{j,i} \forall{i} \in \mathcal{C}, \forall{j} \in \mathcal{K}_i.
\end{equation}

The signal-to-interference and noise ratio (SINR) at which $u_{j,i}$ decodes the message of weaker users $u_{d,i}, \forall{d} \in \{j+1,j+2,...,K_i\}$ can be written as
\begin{align}
SINR_{j,i}^d=\frac{|h_{j,i}|^2p_{d,i}^2}{|h_{j,i}|^2\sum_{s=1}^{d-1}p_{s,i}^2+\sigma_{j,i}^2}, \notag \\ 
\forall{i} \in \mathcal{C}, \forall{j} \in \mathcal{K}_i, \forall{d} \in \{j+1,j+2,...,K_i\}.
\end{align} 

In particular, the strong user $u_{j,i}$ has the capability to perform SIC if and only if the messages intended to the weaker users, i.e., $u_{d,i}$, $\forall{d} \in \{j+1,j+2,...,K_i\}$, are received at $u_{j,i}$ with higher SINR compared to that of the users with stronger channel conditions \cite{Kan}. In fact, this can be only achieved by including the following constraint in the design \cite{Al-Obiedollah}:
\begin{equation} 
p^2_{K,i} \geq p^2_{{K-1},i} \geq ... \geq p^2_{1,i},\forall{i} \in \mathcal{C}.
\end{equation}
The above constraint in (6) is referred to as the SIC constraint in this paper. Therefore, the SINR of $u_{j,i}$ can be defined as \cite{Hanif}
\begin{align}
SINR_{j,i}=\min\{SINR_{j,i}^1, SINR_{j,i}^2,..., SINR_{j,i}^j\}, \notag \\
\forall{i} \in \mathcal{C}, \forall{j} \in \mathcal{K}_i.
\end{align}
Accordingly, based on the above definition, the achieved rate at $u_{j,i}$ can be written as follows:
\begin{equation} 
R_{j,i}=t_i\log_2(1+SINR_{j,i}),\forall{i} \in \mathcal{C}, \forall{j} \in \mathcal{K}_i.
\end{equation}
The total required transmit power at the BS is $\sum_{i=1}^C\sum_{j=1}^{K_i}p_{j,i}^2$, which is limited by an upper bound of $P^{max}$. The total transmit power constraint is defined as,
\begin{equation}
P_t=\sum_{i=1}^C\sum_{j=1}^{K_i}p_{j,i}^2 \leq P^{max} .
\end{equation}

\subsection{Problem Formulation}

As we have mentioned in the introduction, user-fairness is one of the crucial requirements for 5G and beyond wireless networks. Hence, we aim to allocate the available resources, i.e., time and transmit power, among the users to maintain a user fairness in terms of the achieved rates in the hybrid TDMA-NOMA system. In particular, the objective is to maximize the minimum achieved rate at the individual users. This can be accomplished by solving the following max-min optimization problem:
\begin{align}
& \textbf{(P1)}:&  &\max_{{p_{j,i}},{t_i}}&  &\min_{1 \leq j \leq K_i,1 \leq i \leq C } R_{j,i}&\\
&& &\textup{s.t.}\quad& &\textup{C1:} \quad \sum_{i=1}^Ct_i \leq T,&\\
&& && &\textup{C2:} \quad P_t \leq P^{max},&\\
&& && &\textup{C3:} \quad p^2_{K,i} \geq p^2_{{K-1},i} \geq ... \geq p^2_{1,i},\forall{i} \in \mathcal{C},&
\end{align}
where the constraint in (11) ensures that the total allocated time does not exceed the available time $T$ for transmission. Furthermore, the constraint in (13) facilitates the successful implementation of SIC. However, additional complexity is introduced due to the joint allocation of both the time and the transmit power to all served users in the system. Furthermore, the non-convex objective function makes the original problem defined in  (10) - (13) more challenging to solve. Hence, we develop an iterative algorithm to realize the solution in the next section. 

\section{Proposed Methodology And Solution}

In this section, we firstly discuss the grouping strategy that is considered to group the users into a number of clusters in the hybrid TDMA-NOMA system. Next, we use approximation techniques to transform the non-convex optimization problem P1 into an approximated convex problem.

The grouping strategy plays a vital role in the performance of hybrid TDMA-NOMA system. In particular, the optimal user-grouping can be determined through an exhaustive search among all possible user-groupings, which is not possible in practical scenarios due to its computational complexity and the limitations on the processing capabilities of practical systems. SIC can be implemented successfully when the difference between the channel strengths of users is high \cite{B}. Therefore, we assume that the total number of users is even and develop a clustering algorithm that takes into account the difference between the channel strengths of the users. Then, the grouping strategy based on the difference of the channel strengths can be defined as
\begin{align}
\left(\{u_{1,1},u_{2,1}\},\{u_{1,2},u_{2,2}\},...,\{u_{1,C},u_{2,C}\}\right) \equiv \notag\\
\left(\{u_{1},u_{K}\},\{u_{2},u_{K-1}\},...,\{u_{\frac{K}{2}},u_{\frac{K}{2}+1}\}\right).
\end{align}

The objective function of problem P1 is non-convex; hence, we first introduce a new slack variable $\gamma$ to approximate it into a convex one. Based on this slack variable, problem P1 can be reformulated equivalently as, 
\begin{align}
& \textbf{(P2)}:&  &\max_{{p_{j,i}},{t_i},\gamma}&  &\gamma& \\
&& &\textup{s.t.}& &\textup{C1:} \quad \sum_{i=1}^Ct_i \leq T,&\\
&& && &\textup{C2:} \quad P_t \leq P^{max}, &\\
&& && &\textup{C3:} \quad p^2_{K,i} \geq p^2_{{K-1},i} \geq ... \geq p^2_{1,i},\forall{i} \in \mathcal{C}, &\\
&& && &\textup{C4:} \quad t_i\log_2(1+SINR_{j,i}) \geq \gamma, \notag \\
&& && & \quad \quad \quad \forall{i} \in \mathcal{C}, \forall{j} \in \mathcal{K}_i. & 
\end{align}
Note that the objective function of the original optimization problem P1 is replaced with a new single scalar slack variable by using epigraph. However, this non-convex objective function has been formulated into a constraint in (19) (i.e., C4) in P2. In other words, maximizing $\min_{1 \leq j \leq K_i,1 \leq i \leq C } R_{j,i}$ is equivalent to maximizing the slack variable $\gamma$ with a new constraint C4. However, the overall problem still remains intractable due to the non-convex constraints C3 and C4 in P2. In order to solve this non-convex problem, we exploit the SCA technique, in which a set of lower bounded convex terms are introduced to approximate the non-convex terms in the constraints C3 and C4 \cite{H}.

We start handling the non-convexity of the constraint C4 by introducing new slack variables $\alpha_{j,i}$ and $\vartheta_{j,i}$, such that
\begin{small}
\begin{align}
(1+SINR_{j,i}^d) \geq \alpha_{j,i},\forall{i} \in \mathcal{C}, \forall{j} \in \mathcal{K}_i,\forall{d} \in \{j+1,...,K_i\},\\
\log_2(1+SINR_{j,i}) \geq \vartheta_{j,i},\forall{i} \in \mathcal{C},\forall{j} \in \mathcal{K}_i,\\
\alpha_{j,i} \geq 2^{\vartheta_{j,i}},\forall{i} \in \mathcal{C}, \forall{j} \in \mathcal{K}_i,\\
t_i\vartheta_{j,i} \geq \gamma, \forall{i} \in \mathcal{C},\forall{j} \in \mathcal{K}_i,  
\end{align}
\end{small}where the constraint in (22) is convex. Next, to address the non-convexity issue of the constraint in (20), we firstly introduce another slack variable $\eta_{j,i}$, such that
\begin{align}
\frac{|h_{j,i}|^2p_{d,i}^2}{|h_{j,i}|^2\sum_{s=1}^{d-1}p_{s,i}^2+\sigma_{j,i}^2} \geq \frac{(\alpha_{j,i}-1){\eta^2_{j,i}}}{{\eta^2_{j,i}}},\notag\\
\forall{i} \in \mathcal{C}, \forall{j} \in \mathcal{K}_i, \forall{d} \in \{j+1,j+2,...,K_i\}.
\end{align}
Secondly, we decompose the above constraint in (24) into two constraints as follows:
\begin{small}
\begin{equation}
{|h_{j,i}|^2p_{d,i}^2} \geq (\alpha_{j,i}-1){\eta^2_{j,i}} ,
\forall{i} \!\in\! \mathcal{C}, \forall{j} \!\in\! \mathcal{K}_i,\forall{d} \!\in\! \{j+1,j+2,...,K_i\},
\end{equation}
\end{small}
\begin{small}
\begin{equation}
{|h_{j,i}|^2\sum_{s=1}^{d-1}p_{s,i}^2+\sigma_{j,i}^2} \leq {{\eta^2_{j,i}}},
\forall{i} \!\in\! \mathcal{C}, \forall{j} \!\in\! \mathcal{K}_i,\forall{d} \!\in\! \{j+1,j+2,...,K_i\}.
\end{equation}
\end{small}
Then, we exploit the first-order Taylor series to approximate both sides of (25) with  linear convex terms, such that
\begin{align}
&|h_{j,i}|^2\left({p^2_{d,i}}^{(t)}+2p^{(t)}_{d,i}(p_{d,i}-p^{(t)}_{d,i})\right) \geq {{\eta^2_{j,i}}}^{(t)}\left({\alpha}^{(t)}_{j,i}-1\right)\notag\\
&+2\left({\alpha}^{(t)}_{j,i}-1\right){\eta}^{(t)}_{j,i}\left({\eta_{j,i}}-{\eta_{j,i}}^{(t)}\right)+{{\eta_{j,i}}^2}^{(t)}\left({\alpha_{j,i}}-{\alpha_{j,i}}^{(t)}\right),\notag\\
&~~~~~~~~~~~~~\forall{i} \in \mathcal{C}, \forall{j} \in \mathcal{K}_i, \forall{d} \in \{j+1,j+2,...,K_i\},
\end{align}
where $p^{(t)}_{d,i}$, ${\eta_{j,i}}^{(t)}$ and ${\alpha}^{(t)}_{j,i}$ represent the approximations of $p_{d,i}$, ${\eta_{j,i}}$ and ${\alpha}_{j,i}$ at the $t$th iteration, respectively. The constraint in (26) can be rewritten as follows using SOC \cite{boyd,Luo}:
\begin{align}
\Big\lVert{{|h_{j,i}|p_{1,i}, |h_{j,i}|p_{2,i},..., |h_{j,i}|p_{{d-1},i}}, {\sigma_{j,i}}}\Big\lVert \leq \eta_{j,i}, \notag\\
\forall{i} \in \mathcal{C}, \forall{j} \in \mathcal{K}_i,\forall{d} \in \{j+1,j+2,...,K_i\}.
\end{align}

Based on the above multiple slack variables $\alpha_{j,i}$ and $\eta_{j,i}$, the constraint in (20) can be approximated with the convex constraints in (27) and (28).

We address the non-convexity issue of the constraint in (23) by formulating it as the following SOC constraint \cite{boyd,Luo}:
\begin{align}
t_i+\vartheta_{j,i} \geq \Big\lVert{{2\sqrt{\gamma}} \atop {t_i-\vartheta_{j,i}}}\Big\lVert, \forall{i} \in \mathcal{C}, \forall{j} \in \mathcal{K}_i.
\end{align}
Furthermore, each non-convex term in C3 can be approximated by a lower bounded convex term using the first-order Taylor series expansion,
\begin{align}
p^2_{K,i} \geq {p^2_{K,i}} ^{(t)}+2{p_{K,i}}^{(t)}(p_{K,i}-{p_{K,i}}^{(t)}), \forall{i} \in \mathcal{C}.
\end{align}

Therefore, the original non-convex optimization problem P1 can be approximated by the following convex optimization problem:
\begin{align}
& \textbf{(P3)}:&  &\max_{\Gamma}&  \gamma& \\
&& &\textup{s.t.}& &\textup{C1:} \quad \sum_{i=1}^Ct_i \leq T, &\\
&& && &\textup{C2:} \quad \sum_{i=1}^C\sum_{j=1}^{K_i}p_{j,i}^2 \leq P^{max},  &\\
&& && &\textup{C3:} \quad (18),(22),(27),(28),(29),&
\end{align}
where $\Gamma$ consists of all the optimization variables, such that $\Gamma=\{{p_{j,i}},{t_i},\gamma,\alpha_{j,i},\vartheta_{j,i},\eta_{j,i}\}$, $\forall{i} \in \mathcal{C}, \forall{j} \in \mathcal{K}_i$. It is worth pointing out that the solution of P1 is obtained iteratively, such that the approximated convex optimization problem P3 is solved at each iteration. In particular, this requires appropriate selection of the initial variables, i.e., $\Gamma^{(0)}$. These initial values can be chosen by defining random power allocations $p_{j,i}^{(0)}$ that fulfills the maximum power constraint in (9). Then, the corresponding slack variables are evaluated by substituting these power allocations in (27) and (28). The solutions obtained in each iteration are used as initial points for the Taylor series approximation to the next iteration. In fact, the iterative algorithm keeps improving the solutions at each iteration until the difference between two consecutive objective values is less than a pre-defined threshold, $\epsilon$, (i.e., $|\gamma^{*(n+1)}-\gamma^{*(n)}| \leq \epsilon $). The proposed iterative algorithm is summarized in Table I.

\begin{table} 
	\caption{Max-min Joint Resource Allocation Algorithm.}
	\vspace{-4mm}
	\small\noindent \rule[0 ex]{0.49\textwidth}{1.0pt} \\
    \textbf{Algorithm:} Solving Max-min Joint Resource Allocation Problem.  \\
    \noindent \rule[1 ex]{0.49\textwidth}{1.0pt} \\
    1:\hspace*{0.5 mm}\noindent Group the users into clusters based on the grouping strategy \\
    \hspace*{2 mm} defined in (14), \\
    2:\hspace*{0.5 mm}\noindent Initialize: Set the parameters $\Gamma^{(0)}$, \\
    3:\hspace*{0.5 mm}\noindent Repeat \\
    4:\hspace*{3 mm} $\big|$ Solve the problem {P3} in (31) - (34), \\
    5:\hspace*{3 mm} $\big|$ Update all parameters ${\Gamma^{(n)}}$ based on (18),(22),(27),(28),(29), \\
    6:\hspace*{0.5 mm}\noindent {Until} $|\gamma^{*(n+1)}\!-\!\gamma^{*(n)}| \leq    \epsilon$, where $\epsilon$ is a predefined error tolerance \\
    \hspace*{2 mm} threshold. \\
    \small\noindent \rule[0 ex]{0.49\textwidth}{1.0pt} \\
    \vspace{-10mm}  
\end{table}

\section{Simulation Results}

To evaluate the performance of the proposed hybrid TDMA-NOMA scheme with opportunistic time allocations, we compare its performance with that of the conventional scheme with equal time allocations. In simulations, the users are uniformly distributed over a circle with a radius of 50 meters around the BS, and the minimum distance is selected such that $d_0=1$ meters. The corresponding channel gain is $|h_{j,i}|^2=\frac{\beta}{{(d_{j,i}/d_0)}^{\kappa}}$, where $d_{j,i}$ is the distance between $u_{j,i}$ and the BS, measured in meters and $\beta=-30$ dB is the signal attenuation at $d_0$, and $\kappa=2$ is the path-loss exponent. We consider ten users ($K = 10$) divided into five clusters ($C=5$); accordingly, each cluster consists of two users. The noise variance at each user is assumed to be -100 dBm/Hz ($\sigma_{j,i}^2=-100$). Furthermore, we use the CVX software to solve the convex problems in this work \cite{cvx}.

Now, we compare the performance of the hybrid TDMA-NOMA with the opportunistic time allocations versus that of the conventional schemes with equal time allocation in Table II and III. As seen in Table II, the proposed opportunistic time allocations based hybrid TDMA-NOMA outperforms the conventional scheme with equal time allocations in terms of minimum achieved rate. In particular, the opportunistic time allocations provides improvement to the overall system performance. Furthermore, Table III presents the power allocations of all users in the system for both schemes: with opportunistic and equal time allocations. 

Next, Fig. 3 depicts the performance of these schemes in terms of the minimum achieved rate for different transmission power $P^{max}$. Simulation results confirm that the proposed scheme with opportunistic time allocation outperforms the scheme with equal time allocation in terms of the minimum achieved rate. 

Finally, we present simulation results to validate the convergence of the proposed algorithm in Fig. 4. Five different channels are considered, and as seen in Fig. 4, the proposed algorithm converges within a few number of iterations.

\begin{table*} 
	\caption{Power Allocations For Each User In The Hybrid TDMA-NOMA Through The Proposed Opportunistic Time Allocations And The Conventional Equal Time One.}
	\vspace{-4mm}
   \begin{center}	
    \renewcommand\arraystretch{1.4}   
    \resizebox{\textwidth}{!}{
    \begin{tabular}{|c|c|c|c|c|c|c|c|c|c|c|c|c|c|c|c|c|c|c|c|}
    \hline
    \multicolumn{10}{|c|}{Scheme with opportunistic time allocations}&\multicolumn{10}{|c|}{Scheme with equal time allocations} \\  
   	\hline	
    {$p_{1,1}$}&{$p_{2,1}$}&{$p_{1,2}$}&{$p_{2,2}$}&{$p_{1,3}$}&{$p_{2,3}$}&{$p_{1,4}$}&{$p_{2,4}$}&{$p_{1,5}$}&{$p_{2,5}$}&{$p_{1,1}$}&{$p_{2,1}$}&{$p_{1,2}$}&{$p_{2,2}$}&{$p_{1,3}$}&{$p_{2,3}$}&{$p_{1,4}$}&{$p_{2,4}$}&{$p_{1,5}$}&{$p_{2,5}$}\\
    \hline	
    {0.081}&{1.483}&{0.098}&{1.462}&{0.122}&{1.414}&{0.179}&{1.330}&{0.256}&{1.330}&{0.057}&{1.407}&{0.071}&{1.407}&{0.092}&{1.407}&{0.141}&{1.407}&{0.264}&{1.407}\\
    \hline	
    {0.120}&{1.409}&{0.178}&{1.391}&{0.266}&{1.391}&{0.273}&{1.391}&{0.301}&{1.391}&{0.079}&{1.397}&{0.126}&{1.397}&{0.243}&{1.397}&{0.253}&{1.397}&{0.302}&{1.397}\\
   	\hline
   	{0.144}&{1.386}&{0.188}&{1.386}&{0.243}&{1.386}&{0.370}&{1.386}&{0.383}&{1.386}&{0.089}&{1.384}&{0.125}&{1.384}&{0.177}&{1.384}&{0.420}&{1.384}&{0.439}&{1.384}\\
   	\hline
   	{0.034}&{1.572}&{0.036}&{1.523}&{0.091}&{1.306}&{0.121}&{1.306}&{0.253}&{1.306}&{0.023}&{1.409}&{0.023}&{1.409}&{0.053}&{1.409}&{0.079}&{1.409}&{0.263}&{1.409}\\
   	\hline
   	{0.074}&{1.410}&{0.105}&{1.410}&{0.123}&{1.410}&{0.127}&{1.410}&{0.160}&{1.410}&{0.040}&{1.410}&{0.072}&{1.410}&{0.097}&{1.410}&{0.102}&{1.410}&{0.160}&{1.410}\\
   	\hline
   	
    \end{tabular} }		
    \end{center}
\end{table*}

\begin{table*}
	\caption{ Time Allocation And Achieved Minimum Throughout In The Hybrid TDMA-NOMA Through The Proposed Opportunistic Time Allocations And The Conventional Equal Time One.}
   \begin{center}	
    \renewcommand\arraystretch{1.4}
    \begin{tabular}{|c|c|c|c|c|c|c|c|c|c|c|c|c|}
    \hline
    {}&\multicolumn{6}{|c|}{Scheme with opportunistic time allocations}&\multicolumn{6}{|c|}{Scheme with equal time allocations} \\  
   	\hline	
   	{Channels}&{$t_1(s)$}&{$t_2(s)$}&{$t_3(s)$}&{$t_4(s)$}&{$t_5(s)$}&{Minimum throughput}&{$t_1(s)$}&{$t_2(s)$}&{$t_3(s)$}&{$t_4(s)$}&{$t_5(s)$}&{Minimum throughput}\\
   	{}&{}&{}&{}&{}&{}&{(bit/second)}&{}&{}&{}&{}&{}&{(bit/second)}\\
   	\hline	
   	{Channel 1}&{1.878}&{1.904}&{1.934}&{1.980}&{2.304}&{10.664}&{2}&{2}&{2}&{2}&{2}&{9.423}\\
   	\hline
   	{Channel 2}&{1.751}&{1.819}&{2.097}&{2.114}&{2.219}&{9.706}&{2}&{2}&{2}&{2}&{2}&{8.765}\\
	\hline
	{Channel 3}&{1.714}&{1.789}&{1.886}&{2.284}&{2.327}&{8.433}&{2}&{2}&{2}&{2}&{2}&{7.211}\\
	\hline
	{Channel 4}&{1.948}&{1.896}&{1.817}&{1.908}&{2.432}&{11.494}&{2}&{2}&{2}&{2}&{2}&{9.667}\\
	\hline
	{Channel 5}&{1.744}&{1.928}&{2.031}&{2.044}&{2.253}&{14.124}&{2}&{2}&{2}&{2}&{2}&{12.540}\\
	\hline
	
    \end{tabular} 		
    \end{center}
    \vspace{-4mm}
\end{table*}

\begin{figure}
\centering
\includegraphics[width=0.55\textwidth]{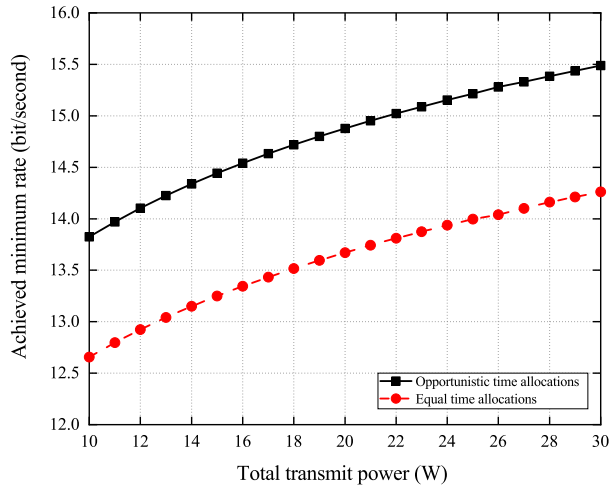}
\caption{The achieved minimum rate versus different total transmit power.}
\label{fig_graph}
\vspace{-4mm}
\end{figure}

\begin{figure}
\centering
\includegraphics[width=0.55\textwidth]{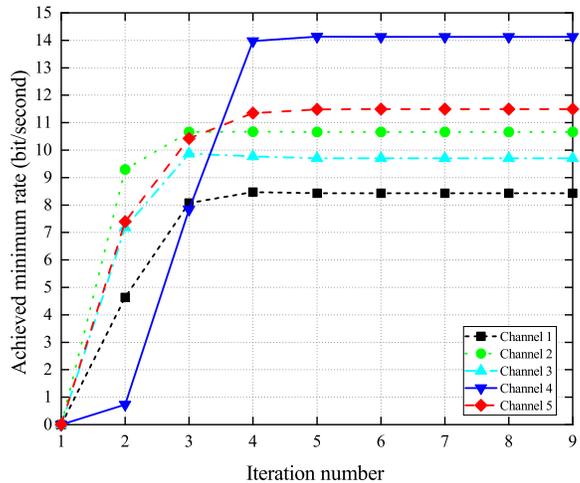}
\caption{The convergence of the SCA algorithm for five different channels.}
\label{fig_graph}
\vspace{-4mm}
\end{figure}

\section{Conclusion}
In this paper, we have studied the max-min joint resource allocation problem of a SISO hybrid TDMA-NOMA system. Specifically, the available transmission time is divided into several time-slots and power-domain NOMA is exploited to serve multiple users within a cluster. However, the formulated max-min optimization problem is non-convex in nature. To cope with this challenge, we develop an iterative algorithm by exploiting SCA and a novel form of a SOC formulation to realize a solution to the original problem. Simulation results demonstrate that the proposed hybrid TDMA-NOMA system outperforms the conventional resource allocations with equal time assignment in terms of the minimum achieved rate and overall system throughput.

\balance
%----------------------------------------------------------------------------------------------------------------------------


\begin{thebibliography}{99}

\bibitem{Y} Y. Liu, Z. Qin, M. Elkashlan, Z. Ding, A. Nallanathan, and L. Hanzo, ``Non-orthogonal multiple access for 5G and beyond,'' {\em Proceedings of the IEEE}, vol. 105, no. 12, pp. 2347-2381, December 2017.

\bibitem{alavi2} F. Alavi, K. Cumanan, Z. Ding, and A. G. Burr, ``Robust beamforming techniques for non-orthogonal multiple access systems with bounded channel uncertainties,'' {\em IEEE Communications Letters}, vol. 21, no. 9, pp. 2033-2036, September 2017.

\bibitem{Benjebbour} A. Benjebbour, A. Li, K. Saito, Y. Saito, Y. Kishiyama, and T. Nakamura, ``NOMA: From concept to standardization,'' in~{\em Proc. IEEE Conference on Standards for Communications and Networking (CSCN)}, Tokyo, Japan, October, 2015, pp. 18-23.

\bibitem{Demestichas} P. Demestichas, A. Georgakopoulos, D. Karvounas, K. Tsagkaris, V. Stavroulaki, J. Lu, C. Xiong, and J. Yao, ``5G on the horizon: Key challenges for the radio-access network,'' {\em IEEE Vehicular Technology Magazine}, vol. 8, no. 3, pp. 47-53, September 2013.

\bibitem{Alavi} F. Alavi, K. Cumanan, M. Fozooni, Z. Ding, S. Lambotharan, and O. A. Dobre,  ``Robust energy-efficient design for MISO non-orthogonal multiple access systems,'' {\em IEEE Transactions on Communications}, vol. 67, no. 11, pp.7937-7949, November 2019.


\bibitem{Rappaport} T. S. Rappaport, S. Sun, R. Mayzus, H. Zhao, Y. Azar, K. Wang, G. N. Wong, J. K. Schulz, M. Samimi, and F. Gutierrez, ``Millimeter wave mobile communications for 5G cellular: It will work!,'' {\em IEEE Access}, vol. 1, pp. 335-349, May 2013.

\bibitem{Islam} S. R. Islam, N. Avazov, O. A. Dobre, and K. S. Kwak, ``Power-domain non-orthogonal multiple access (NOMA) in 5G systems: Potentials and challenges,'' {\em IEEE Communications Surveys and Tutorials}, vol. 19, no. 2, pp. 721-742, Second Quarter 2017.

\bibitem{Higuchi} K. Higuchi and A. Benjebbour, ``Non-orthogonal multiple access (NOMA) with successive interference cancellation for future radio access,'' {\em IEICE Transactions on Communications}, vol. 98, no. 3, pp. 403-414, March 2015. 

\bibitem{10} P. Xu and K. Cumanan,``Optimal power allocation scheme for non-orthogonal multiple access with $\alpha$-fairness,''{\em IEEE Journal on Selected Areas in Communications}, vol. 35, no. 10, pp.2357-2369, October 2017.

\bibitem{11} J. Luo, J. Tang, D. K. So, G. Chen, K. Cumanan, and J. A. Chambers , ``A deep learning-based approach to power minimization in multi-carrier NOMA with SWIPT,'' {\em IEEE Access}, vol. 7, pp 17450-17460, January 2019.

\bibitem{phd} H. M. Al-Obiedollah, ``Resource allocation techniques for non-orthogonal multiple access systems,'' PhD Thesis, University of York. 2019.

\bibitem{alavi1} F. Alavi, K. Cumanan, Z. Ding, and A. G. Burr, `` Beamforming techniques for non-orthogonal multiple access in 5G cellular networks,'' {\em IEEE Transactions on Vehicular Technology}, vol. 67, no. 10, pp. 9474-9487, October 2018. 


\bibitem{Kan} K. Cumanan, Z. Ding, B. Sharif, G. Y. Tian, and K. K. Leung, ``Secrecy rate optimizations for a MIMO secrecy channel with a multiple-antenna eavesdropper,'' {\em  IEEE Transactions on Vehicular Technology}, vol. 64, no. 4, pp. 1678-1690, May 2014.

\bibitem{Timotheou} S. Timotheou and I. Krikidis, ``Fairness for non-orthogonal multiple access in 5G systems,'' {\em IEEE Signal Processing Letters}, vol. 22, no. 10, pp. 1647-1651, October 2015.

\bibitem{Jingjing} J. Cui, Z. Ding, and P. Fan, ``Power minimization strategies in downlink MIMO-NOMA systems,'' in~{\em Proc. IEEE International Conference on Communications (ICC)}, Paris, France, May, 2017, pp. 1-6.

\bibitem{O.A.Dobre} S. R. Islam, M. Zeng, O. A. Dobre, and K. S. Kwak, ``Resource allocation for downlink NOMA systems: key techniques and open issues,'' {\em  IEEE Wireless Communications}, vol. 25, no. 2, pp. 40-47, April 2018.

\bibitem{M} M. Zeng, A. Yadav, O. A. Dobre, and H. V. Poor, ``Energy-efficient joint User-RB association and power allocation for uplink hybrid NOMA-OMA,'' {\em  IEEE Internet of Things Journal}, vol. 6, no. 3, pp. 5119-5131, June 2019.


\bibitem{H} H. M. Al-Obiedollah, K. Cumanan, A. G. Burr, J. Tang, Y. Rahulamathavan,
Z. Ding, and O. A. Dobre, ``On energy harvesting of hybrid TDMA-NOMA systems,'' arXiv preprint arXiv:1908.08719, 2019.

\bibitem{Elkashlan} Z. Ding, Y. Liu, J. Choi, Q. Sun, M. Elkashlan, and H. V. Poor, ``Application of non-orthogonal multiple access in LTE and 5G networks,'' {\em IEEE Communication Magazine}, vol. 55, no. 2, pp. 185-191, February 2017.

\bibitem{Fang} F. Fang, H. Zhang, J. Cheng, and V. C. Leung, ``Energy-efficient
resource allocation for downlink non-orthogonal multiple access network,'' {\em
IEEE Transactions on Communications}, vol. 64, no. 9, pp. 3722-3732, September 2016.

\bibitem{Wei} Z. Wei, D. W. K. Ng, and J. Yuan, ``Power-efficient resource allocation 
for MC-NOMA with statistical channel state information,'' in~{\em Proc. IEEE Global Communications Conference (GLOBECOM)}, Washington, DC, USA, December, 2016, pp. 1-7. 

\bibitem{Wei2} Z. Wei, D. W. K. Ng, J. Yuan, and H. M. Wang, ``Optimal resource allocation for power-efficient MC-NOMA with imperfect channel state information,'' {\em IEEE Transactions on Communications}, vol. 65, no. 9, pp. 3944-3961, September 2017.

\bibitem{Diamantoulakis} A. J. Muhammed, Z. Ma, P. D. Diamantoulakis, L. Li, and G. K. Karagiannidis, ``Energy-efficient resource allocation in multicarrier noma systems with fairness,'' {\em IEEE Transactions on Communications}, vol. PP, pp. 1-1, September 2019.

\bibitem{Beck} A. Beck, A. Ben-Tal, and L. Tetruashvili, ``A sequential parametric convex approximation method with applications to nonconvex truss topology design problems,'' {\em Journal of Global Optimization}, vol. 47, no. 1, pp. 29-51, May 2010.

\bibitem{xinchen}K. Chi, X. Wei, Y. Li, and X. Tian, ``Throughput maximization in wireless powered communication networks with minimum node throughput requirement,'' {\em International Journal of Communication Systems}, vol. 31, no. 15, p.e 3775, September 2018.

\bibitem{Al-Obiedollah} H. M. Al-Obiedollah, K. Cumanan, J. Thiyagalingam, A. G. Burr,  Z. Ding, and O. A. Dobre, ``Energy efficient beamforming design for MISO non-orthogonal multiple access systems,'' {\em IEEE Transactions on Communications}, vol. 67, no. 6, pp. 4117-4131, June 2019.

\bibitem{Hanif} M. F. Hanif, Z. Ding, T. Ratnarajah, and G. K. Karagiannidis, ``A minorization-maximization method for optimizing sum rate in the downlink of non-orthogonal multiple access systems,'' {\em IEEE Transactions on Signal Processing}, vol. 64, no. 1, pp. 76-88, January 2016.

\bibitem{B} B. Kim, S. Lim, H. Kim, S. Suh, J. Kwun, S. Choi, C. Lee, S. Lee, and D. Hong, ``Non-orthogonal multiple access in a downlink multiuser beamforming system,'' in~{\em Proc. IEEE Military Communications Conference (MILCOM)}, San Diego, CA, USA, November, 2013, pp. 1278-1283.

\bibitem{boyd} S. Boyd and L. Vandenberghe, ``Convex optimization,'' {\em Cambridge University Press}, 2004.

\bibitem{Luo} Z. Q. Luo and W. Yu, ``An introduction to convex optimization for communications and signal processing,'' {\em IEEE Journal on Selected Areas in Communications}, vol. 24, no. 8, pp. 1426-1438, August 2006.

\bibitem{cvx}M. Grant and S. Boyd, CVX: Matlab software for disciplined convex programming. ver.2.1, Feb.2015[Online]. Available: http://cvxr.com/cvx/.




\end{thebibliography}
\end{document}